# A tool for computing diversity and consideration on differences between diversity indices

Ciprian Palaghianu

Abstract
Diversity represents a key concept in ecology, and there are various methods of assessing it. The multitude of diversity indices are quite puzzling and sometimes difficult to compute for a large volume of data. This paper promotes a computational tool used to assess the diversity of different entities. The BIODIV software is a user-friendly tool, developed using Microsoft Visual Basic. It is capable to compute several diversity indices such as: Shannon, Simpson, Pielou, Brillouin, Berger-Parker, McIntosh, Margalef, Menhinick and Gleason. The software tool was tested using real data sets and the results were analysed in order to make assumption on the indices behaviour. The results showed a clear segregation of indices in two major groups with similar expressivity.

Introduction
The keywords such as biodiversity, diversity or heterogeneity are extensively used in ecological studies. So diversity, as a measure of heterogeneity still represents an important concept in ecology despite all the new trends in this field. The pioneer work of Gleason (1922), Shannon (1948), Simpson (1949) or Pielou (1969) was continued later in numerous studies.
There are different aspects of diversity that can be assessed at landscape level, population level or even regarding to certain individual attributes. But why diversity is such an important feature? That is because diversity and heterogeneity of a system are frequently related to its superior stability. A more complex and diverse system can manifest a higher resilience to external disturbances.
In forestry, due to a relatively low number of tree species in the stands located in the temperate region, the researchers rather focus on structural or dimensional diversity. This type of diversity is also related to a higher structural stability and the interest in this particular field was constant in the last decades (Zenner & Hibbs, 2000; Pommerening, 2002; Zenner, 2005; Davies & Pommerening, 2008).
But whatever the theme of the study is, the measurement of diversity still remains a puzzling issue. There is a huge diversity regarding the possibility of assessing diversity. Ecology offers a great variety of techniques and indices for measuring this highly appreciated feature. And this represents in fact a genuine problem, a scientifically dilemma, because it is quite challenging to decide which method or index is more suited to use in your research.
Although there are several comprehensive comparative studies that debate on the quality and sensitivity of the main diversity indices (Staudhammer & LeMay, 2001; Pommerening, 2006; Lexerod & Eid, 2006), the verdict is still unclear, and it was not established the superiority of one particular index.
This paper is not trying either to set a verdict, but it might cast some light on various aspects regarding computing and interpreting indices values.

Material and methods
The computational process required by the diversity indices is rather complex and difficult, especially for large amount of data. So the need for a computer becomes evident. The advancement in spreadsheets has made this procedure more approachable, but even with the help of such software (e.g. Microsoft Excel) several numerical operations require further knowledge of VBA coding (Visual Basic for Applications). This fact might become limitative for some researchers, so I have developed a computational tool used to assess the diversity. BIODIV software is a standalone program which was coded using Microsoft Visual Basic and it's based on earlier personal studies (Palaghianu & Avăcăriţei, 2006). BIODIV has a graphic user-friendly interface and uses the input data from an MS Excel worksheet (Figure 1).

Using BIODIV, the productivity of computation increase, the researchers have a convenient method of data input and the results are rapidly and easy obtained and saved in a spreadsheet.
The input data consists in classes and their frequency. Using this data, the last version of the software is capable to compute several diversity indices such as: Shannon, Simpson, Pielou, Brillouin, Berger-Parker, McIntosh, Margalef, Menhinick and Gleason.

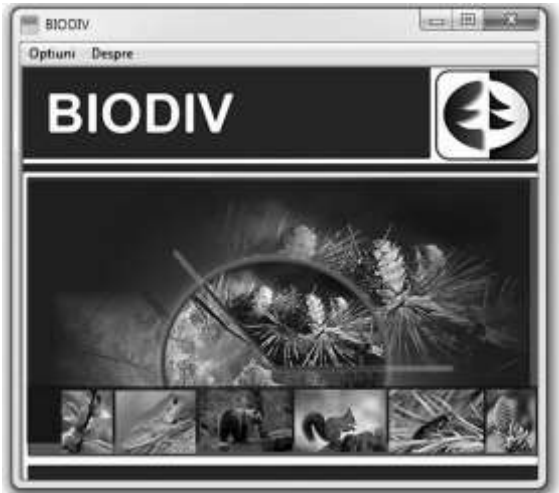

Fig. 1: The BIODIV software interface

a) The Shannon index (H)
This is one of the most frequently used biodiversity index. It originates from the information theory as a measure of entropy (Shannon, 1948) and sometimes it is incorrectly mentioned as Shannon-Wiener or Shannon-Weaver index.

$$H = -\sum_{i=1}^{S} p_i \ln p_i, \quad p_i = n_i / N$$

where the same notations are used for all the following expressions:
H - is the value of Shannon index

$p_i$ - is the proportion of each class

$n_i$ - the frequency for the class i
S - the total number of classes
N - the total number of observations

The minimum value of the index is 0 when all the observations belong to a single class. The maximum value equals *ln (1/S)* and it can be reached when the observations are equally divided between all the classes.

b) Evenness or Pielou Index (E)
This index represents a standardization form of the Shannon index, displaying the relations between the class frequencies (Pielou, 1969). The evenness equals one when the class frequencies are similar and it tends to zero when the majority of observations belong to a single class.

$$E = \frac{H}{\ln(S)}$$

c) Brillouin index (HB)
Generally, the value of this index is relatively closed to Shannon index, but it is always lower (Magurran, 2004). From the mathematical perspective, this index is superior in sensitivity to Shannon and that is why it is recommended by many. But the complex and intricate formulae as well as the unexpected results biased by the observation volume discourage the researchers in using it.

$$HB = \frac{\ln(N!) - \sum_{i=1}^{S} \ln(n_i!)}{N}$$

d) The Simpson index (D) is a widely used index that takes into account not only the number of classes, but also the proportion of each class (Simpson, 1949). In general, there are three alternatives of these indices (D, 1-D and 1/D).

$$D = \sum p_i^2, \quad p_i = n_i / N,$$

In this version, the Simpson index (D) represents the possibility that two randomly observations belong to the same class. The minimum value is 1/S (where S is the total number of classes) and the maximum is 1.
The diversity Simpson index (1 – D) represents the possibility that two randomly observations belong to different classes. The minimum value is 0 and the maximum is 1-(1/S).
The Simpson reciprocal index (1/D) expresses the number of classes with a high weight which leads to a specific value of Simpson index D. The minimum is 1 and the maximum reach S value.

e) Berger-Parker index (d)
This index simplifies the diversity assessing, using as reference the dominance or the maximum proportion of a class (Berger & Parker, 1970). Its value does not take into account the number of classes but it is highly influenced by the equity. The minimum value is 1/S in case of a uniform distribution of observations between classes and the maximum extent to 1 for grouping of observations to a single class.

$$d = p_{max} (\forall i : p_{max} \geq p_i)$$

f) McIntosh index ($D_{MI}$) is another evaluation form of dominance (McIntosh, 1967), but the index is rather infrequently used due to its computing complexity. Furthermore its ecological interpretation is controversial.

$$D_{MI} = \frac{N - \sqrt{\sum_{i=1}^{S} n_i^2}}{N - \sqrt{N}}$$

g) Margalef index ($D_{Mg}$) is clearly inspired by the Gleason coefficient (Margalef, 1958) and it is commonly used due to its simplicity. The index value is not biased by the class frequencies but depends on the number of classes. The minimum value drops to 0 for the grouping of all observations to a single class.

$$D_{Mg} = \frac{S-1}{\ln(N)}$$

h) Menhinick index ($D_{Mn}$) resulted after a comparative study on diversity indices (Menhinick, 1964). His author developed the index considering the influence on diversity of the analysis scale therefore the index value is not taking into account class frequencies.

$$D_{Mn} = \frac{S}{\sqrt{N}}$$

g) Glisson coefficient ($K_{gl}$) is one of the earliest indices of diversity (Gleason, 1922), consequently influenced many other indices afterward. Its value is less dependent on the analysis scale due to its logarithmic expression.

$K_{gl} = (S– 1) / \log(N)$

The BIODIV software was tested on real data sets. It was analysed the structural diversity of a sapling population from a natural regeneration spot located in Flămânzi Forest District, parcel 50A, Botoşani County, Romania. The species composition consists of 30% sessile oak, 20% oak, 30% common hornbeam, 10% small-leaved linden and 10% common ash. The biometric features (height, diameter, crown insertion height and two crown diameters) of all 7253 saplings and seedlings were collected from a network of ten permanent sampling plots (7 x 7 m).

Results and Discussion
Using BIODIV, all the diversity indices prior mentioned were computed on classes of diameter, height, crown volume and exterior surface of the crown. The evaluation was made separately by species and one evaluation grouped all the saplings, regardless of species, by the four biometric features.
In Table 1 there are presented only the indices values for the whole population of saplings, regardless of species.
Further, a correlation matrix was computed, using all the indices values, in order to establish the indices relationships and similar behaviours (Table 2 and 3). The correlation coefficients have large values, as expected, indicating strong associations between indices and statistically all the relationships can be designated as highly significant (\*\*\*).

Tab. 1: Diversity indices values

| Index / feature | diameter | height | crown volume | crown surface |
|---|---|---|---|---|
| Simpson (D) | 0,240 | 0,157 | 0,974 | 0,785 |
| Simpson (1-D) | 0,760 | 0,843 | 0,026 | 0,215 |
| Simpson (1/D) | 4,164 | 6,351 | 1,027 | 1,273 |
| Shannon (H) | 1,610 | 1,973 | 0,084 | 0,461 |
| Pielou (E) | 0,610 | 0,729 | 0,036 | 0,186 |
| Brillouin (HB) | 1,606 | 1,967 | 0,082 | 0,458 |
| Berger-Parker (d) | 0,358 | 0,199 | 0,987 | 0,881 |
| McIntosh (Dmi) | 0,516 | 0,610 | 0,013 | 0,115 |
| Margalef (DMg) | 1,462 | 1,575 | 1,012 | 1,237 |
| Menhinick (DMn) | 0,164 | 0,176 | 0,117 | 0,141 |
| Gleason (Kgl) | 1,014 | 1,092 | 0,702 | 0,858 |

Tab. 2: Correlations between diversity indices

|  | *1-D* | *H* | *HB* | *d* |
|---|---|---|---|---|
| 1-D | 1 | 0,995 | 0,994 | -0,994 |
| H | 0,995 | 1 | 1,000 | -0,996 |
| HB | 0,994 | 1,000 | 1 | -0,996 |
| d | -0,994 | -0,996 | -0,996 | 1 |
| Dmi | 0,997 | 0,998 | 0,997 | -0,998 |
| DMg | 0,785 | 0,798 | 0,801 | -0,774 |
| DMn | 0,778 | 0,768 | 0,761 | -0,760 |
| Kgl | 0,785 | 0,798 | 0,801 | -0,774 |

Tab. 3: Correlations between diversity indices

|  | *DMi* | *DMg* | *DMn* | *Kgl* |
|---|---|---|---|---|
| 1-D | 0,997 | 0,785 | 0,778 | 0,785 |
| H | 0,998 | 0,798 | 0,768 | 0,798 |
| HB | 0,997 | 0,801 | 0,761 | 0,801 |
| d | -0,998 | -0,774 | -0,760 | -0,774 |
| Dmi | 1 | 0,776 | 0,782 | 0,776 |
| DMg | 0,776 | 1 | 0,573 | 1,000 |
| DMn | 0,782 | 0,573 | 1 | 0,573 |
| Kgl | 0,776 | 1,000 | 0,573 | 1 |

Analysing the mathematical substantiation of the indices, their values for the different features and the correlations between indices, segregation in two main groups was observed. The first group encompass Shannon, Simpson, Brillouin, Berger-Parker and McIntosh indices. Nearly functional relationships between this indices were detected, with high values of the coefficient of correlation (over 0,990 \*\*\*). The relation between Shannon and Brillouin index is quite functional with a correlation value of 1,00 ***, anticipated by their mathematical similarity.

The second group is an atypical one, encompassing Gleason, Margalef and (questionable) Menhinick indices. We record another expected functional relationship between two indices: Gleason and Margalef.

Indices from this cluster are strongly influenced by Gleason (1922) approach therefore they don't take into account the proportion of the classes. Menhinick index strays from both groups, but mathematically it is closer related to the second one, justifying its classification.

Comparing the expressivity of the two groups, the indices from the first group might be consider superior, due to their higher mathematically complexity. Their values are based not only on the number of classes and observation, but also on the class proportions.

The differences between the indices from the first group are not significant considering their expressivity. Although Shannon index has constantly higher absolute values compared with the rest of the indices, this does not imply a greater sensitivity.

Conclusion

I consider that BIODIV software might improve the productivity of diversity analysis and it is quite user-friendly even for the unexperienced users. The software tool was tested on real data and the results revealed interesting differences and similarities in the behaviour of the studied indices. The results showed a clear segregation of indices in two major groups with different expressivity. Generally, the first group has a more complex mathematic foundation and it seems more sensitive in assessing diversity. However there is no justification for using a whole collection of indices, because all of them share a similar responsiveness. It is sufficient to use only one index from the first group, and I would recommend Shannon or Simpson due to their notoriety which increases the possibility to compare the results. The Simpson index, by all its three versions, offers a better flexibility and has even a better ecological interpretation. Nevertheless, the results can be seriously altered by the way classes are formed.

As a final mention, BIODIV is non-commercial software, and it can be used without any restrictions by researchers.

Acknowledgement
This work has been mainly supported by the Romanian National Authority for Scientific Research, UEFISCDI, Grant PNII-PT-PCCA-2011-3.2-1574, no.119/2012 (STROMA).
The views expressed herein are exclusively those of the author.

Author's contact
Eng. Ciprian Palaghianu, Ph.D.,
Department of Silviculture and Environmental Protection, Forestry Faculty,
Stefan cel Mare University of Suceava
Universitatii Street, 13, Suceava, Romania
Phone: (+40) 0745 614 487
Email: cpalaghianu@usv.ro